\documentclass[aps,prl,showpacs,twocolumn,lineno,groupedaddress]{revtex4}  
\ifx\pdftexversion\undefined
  \usepackage[dvips]{graphics}
\else
  \usepackage[pdftex]{graphics}
\fi
\usepackage{graphicx}
\usepackage{dcolumn}   
\usepackage{bm}        
\usepackage{amssymb}   

\providecommand{\tabularnewline}{\\}

\begin{document}

\preprint{FERMILAB-PUB-04/297-E}

\hspace{5.2in} \mbox{FERMILAB-PUB-04/297-E}

\title{A Measurement of the Ratio of Inclusive Cross Sections $\sigma(p\bar{{p}}\rightarrow Z+b{\rm\ jet})/
\sigma(p\bar{{p}}\rightarrow Z+{\rm jet})$ at $\sqrt{{s}}=1.96$ TeV}

%
\author{                                                                      
V.M.~Abazov,$^{33}$                                                           
B.~Abbott,$^{70}$                                                             
M.~Abolins,$^{61}$                                                            
B.S.~Acharya,$^{27}$                                                          
M.~Adams,$^{48}$                                                              
T.~Adams,$^{46}$                                                              
M.~Agelou,$^{17}$                                                             
J.-L.~Agram,$^{18}$                                                           
S.H.~Ahn,$^{29}$                                                              
M.~Ahsan,$^{55}$
% ...
\\                                                                            
\vskip 0.30cm                                                                 
\centerline{(D\O\ Collaboration)}                                             
\vskip 0.30cm                                                                 
}                                                                             
\address{                                                                     
\centerline{$^{1}$Universidad de Buenos Aires, Buenos Aires, Argentina}       
\centerline{$^{2}$LAFEX, Centro Brasileiro de Pesquisas F{\'\i}sicas,         
                  Rio de Janeiro, Brazil}                                     
\centerline{$^{3}$Universidade do Estado do Rio de Janeiro,                   
                  Rio de Janeiro, Brazil}                                     
\centerline{$^{4}$Instituto de F\'{\i}sica Te\'orica, Universidade            
                  Estadual Paulista, S\~ao Paulo, Brazil}                     
\centerline{$^{5}$Simon Fraser University, Burnaby, Canada, University of     
                  Alberta, Edmonton, Canada,}                                 
\centerline{McGill University, Montreal, Canada and York University,          
                  Toronto, Canada}                                            
\centerline{$^{6}$Institute of High Energy Physics, Beijing,                  
                  People's Republic of China}                                 
\centerline{$^{7}$Universidad de los Andes, Bogot\'{a}, Colombia}             
\centerline{$^{8}$Charles University, Center for Particle Physics,            
                  Prague, Czech Republic}                                     
\centerline{$^{9}$Czech Technical University, Prague, Czech Republic}         
\centerline{$^{10}$Institute of Physics, Academy of Sciences, Center          
                  for Particle Physics, Prague, Czech Republic}               
\centerline{$^{11}$Universidad San Francisco de Quito, Quito, Ecuador}        
\centerline{$^{12}$Laboratoire de Physique Corpusculaire, IN2P3-CNRS,         
                 Universit\'e Blaise Pascal, Clermont-Ferrand, France}        
\centerline{$^{13}$Laboratoire de Physique Subatomique et de Cosmologie,      
                  IN2P3-CNRS, Universite de Grenoble 1, Grenoble, France}     
\centerline{$^{14}$CPPM, IN2P3-CNRS, Universit\'e de la M\'editerran\'ee,     
                  Marseille, France}                                          
\centerline{$^{15}$Laboratoire de l'Acc\'el\'erateur Lin\'eaire,              
                  IN2P3-CNRS, Orsay, France}                                  
\centerline{$^{16}$LPNHE, Universit\'es Paris VI and VII, IN2P3-CNRS,         
                  Paris, France}                                              
\centerline{$^{17}$DAPNIA/Service de Physique des Particules, CEA, Saclay,    
                  France}                                                     
\centerline{$^{18}$IReS, IN2P3-CNRS, Universit\'e Louis Pasteur, Strasbourg,  
                  France and Universit\'e de Haute Alsace, Mulhouse, France}  
\centerline{$^{19}$Institut de Physique Nucl\'eaire de Lyon, IN2P3-CNRS,      
                   Universit\'e Claude Bernard, Villeurbanne, France}         
\centerline{$^{20}$RWTH Aachen, III. Physikalisches Institut A,               
                   Aachen, Germany}                                           
\centerline{$^{21}$Universit{\"a}t Bonn, Physikalisches Institut,             
                  Bonn, Germany}                                              
\centerline{$^{22}$Universit{\"a}t Freiburg, Physikalisches Institut,         
                  Freiburg, Germany}                                          
\centerline{$^{23}$Universit{\"a}t Mainz, Institut f{\"u}r Physik,            
                  Mainz, Germany}                                             
\centerline{$^{24}$Ludwig-Maximilians-Universit{\"a}t M{\"u}nchen,            
                   M{\"u}nchen, Germany}                                      
\centerline{$^{25}$Fachbereich Physik, University of Wuppertal,               
                   Wuppertal, Germany}                                        
\centerline{$^{26}$Panjab University, Chandigarh, India}                      
\centerline{$^{27}$Tata Institute of Fundamental Research, Mumbai, India}     
\centerline{$^{28}$University College Dublin, Dublin, Ireland}                
\centerline{$^{29}$Korea Detector Laboratory, Korea University,               
                   Seoul, Korea}                                              
\centerline{$^{30}$CINVESTAV, Mexico City, Mexico}                            
\centerline{$^{31}$FOM-Institute NIKHEF and University of                     
                  Amsterdam/NIKHEF, Amsterdam, The Netherlands}               
\centerline{$^{32}$University of Nijmegen/NIKHEF, Nijmegen, The               
                  Netherlands}                                                
\centerline{$^{33}$Joint Institute for Nuclear Research, Dubna, Russia}       
\centerline{$^{34}$Institute for Theoretical and Experimental Physics,        
                  Moscow, Russia}                                             
\centerline{$^{35}$Moscow State University, Moscow, Russia}                   
\centerline{$^{36}$Institute for High Energy Physics, Protvino, Russia}       
\centerline{$^{37}$Petersburg Nuclear Physics Institute,                      
                   St. Petersburg, Russia}                                    
\centerline{$^{38}$Lund University, Lund, Sweden, Royal Institute of          
                   Technology and Stockholm University, Stockholm,            
                   Sweden and}                                                
\centerline{Uppsala University, Uppsala, Sweden}                              
\centerline{$^{39}$Lancaster University, Lancaster, United Kingdom}           
\centerline{$^{40}$Imperial College, London, United Kingdom}                  
\centerline{$^{41}$University of Manchester, Manchester, United Kingdom}      
\centerline{$^{42}$University of Arizona, Tucson, Arizona 85721, USA}         
\centerline{$^{43}$Lawrence Berkeley National Laboratory and University of    
                  California, Berkeley, California 94720, USA}                
\centerline{$^{44}$California State University, Fresno, California 93740, USA}
\centerline{$^{45}$University of California, Riverside, California 92521, USA}
\centerline{$^{46}$Florida State University, Tallahassee, Florida 32306, USA} 
\centerline{$^{47}$Fermi National Accelerator Laboratory, Batavia,            
                   Illinois 60510, USA}                                       
\centerline{$^{48}$University of Illinois at Chicago, Chicago,                
                   Illinois 60607, USA}                                       
\centerline{$^{49}$Northern Illinois University, DeKalb, Illinois 60115, USA} 
\centerline{$^{50}$Northwestern University, Evanston, Illinois 60208, USA}    
\centerline{$^{51}$Indiana University, Bloomington, Indiana 47405, USA}       
\centerline{$^{52}$University of Notre Dame, Notre Dame, Indiana 46556, USA}  
\centerline{$^{53}$Iowa State University, Ames, Iowa 50011, USA}              
\centerline{$^{54}$University of Kansas, Lawrence, Kansas 66045, USA}         
\centerline{$^{55}$Kansas State University, Manhattan, Kansas 66506, USA}     
\centerline{$^{56}$Louisiana Tech University, Ruston, Louisiana 71272, USA}   
\centerline{$^{57}$University of Maryland, College Park, Maryland 20742, USA} 
\centerline{$^{58}$Boston University, Boston, Massachusetts 02215, USA}       
\centerline{$^{59}$Northeastern University, Boston, Massachusetts 02115, USA} 
\centerline{$^{60}$University of Michigan, Ann Arbor, Michigan 48109, USA}    
\centerline{$^{61}$Michigan State University, East Lansing, Michigan 48824,   
                   USA}                                                       
\centerline{$^{62}$University of Mississippi, University, Mississippi 38677,  
                   USA}                                                       
\centerline{$^{63}$University of Nebraska, Lincoln, Nebraska 68588, USA}      
\centerline{$^{64}$Princeton University, Princeton, New Jersey 08544, USA}    
\centerline{$^{65}$Columbia University, New York, New York 10027, USA}        
\centerline{$^{66}$University of Rochester, Rochester, New York 14627, USA}   
\centerline{$^{67}$State University of New York, Stony Brook,                 
                   New York 11794, USA}                                       
\centerline{$^{68}$Brookhaven National Laboratory, Upton, New York 11973, USA}
\centerline{$^{69}$Langston University, Langston, Oklahoma 73050, USA}        
\centerline{$^{70}$University of Oklahoma, Norman, Oklahoma 73019, USA}       
\centerline{$^{71}$Brown University, Providence, Rhode Island 02912, USA}     
\centerline{$^{72}$University of Texas, Arlington, Texas 76019, USA}          
\centerline{$^{73}$Southern Methodist University, Dallas, Texas 75275, USA}   
\centerline{$^{74}$Rice University, Houston, Texas 77005, USA}                
\centerline{$^{75}$University of Virginia, Charlottesville, Virginia 22901,   
                   USA}                                                       
\centerline{$^{76}$University of Washington, Seattle, Washington 98195, USA}  
}                                                                             

\date{\today}

\begin{abstract}
Using the data collected with the D\O\ detector 
at $\sqrt{{s}}=1.96$ TeV, for integrated luminosities of about $180\ {\rm pb}^{-1}$,
we have measured the ratio of inclusive
cross sections for $p\bar{p}\rightarrow Z+b$ jet to 
$p\bar{{p}}\rightarrow Z{\rm+jet}$ production.
The inclusive $Z+b$-jet reaction is an important background to searches for
the Higgs boson in associated $ZH$ production 
at the Fermilab Tevatron collider. 
Our measurement is the first of its kind, and relies on the $Z\rightarrow e^{+}e^{-}$ 
and $Z\rightarrow\mu^{+}\mu^{-}$ modes. 
The combined measurement of the ratio yields $0.023\pm0.005$ for hadronic jets 
with transverse momenta $p_{T}>20$ GeV/$c$ and pseudorapidities $|\eta|<2.5$, consistent with next-to-leading order predictions
of the standard model.
\end{abstract}
\pacs{14.70.Hp, 14.65.Fy}
\maketitle

Inclusive $Z+b$-jet production is expected to be a major background to Higgs production 
in the $p\bar{p}\rightarrow ZH$ channel, with subsequent Higgs-boson decays into $b\bar{{b}}$. 
The parton-level subprocesses expected to contribute to the $Z+b$-jet final state are
$bg \rightarrow Zb$ (where $g$ stands for a gluon), 
and $q \bar q \rightarrow Zg$, with $g \rightarrow b \bar b$ \cite{campbell}. The process $bg\rightarrow Zb$, where
the initial $b$ is from the sea of the proton parton distribution, is predicted to
account for approximately two thirds 
of the total inclusive cross section $\sigma(p\bar{p}\rightarrow Z+b{\rm\ jet})$ at $\sqrt{s}=1.96$ TeV.
The $b$-quark density of the proton influences the production rates 
of single top quarks and the final state $hb$, with $h$ representing a supersymmetric Higgs boson.
Consequently, the measurement of $Z+b$ jet production is an important step 
in constraining the $b$-quark density of protons.

In this Letter, we describe a measurement of the ratio of production 
cross sections of inclusive $Z+b$ jets to $Z+$jets. 
The measurement of the ratio benefits from cancellations of many systematic uncertainties, such as the 6.5\% uncertainty 
in the luminosity, and therefore allows a more precise comparison with theory.

We search for $Z$ bosons in association with hadronic jets in
about $180\  {\rm pb}^{-1}$ of data collected at the D\O\ experiment
between August 2002 and September 2003.
The D\O\ detector at the Fermilab Tevatron collider is a 
general-purpose detector comprising a magnetic
central-tracking, preshower, calorimeter, and muon systems~\cite{run2det}. 
The central-tracking system consists of a silicon microstrip tracker
(SMT) and a central fiber tracker, both located within a 2~T
superconducting solenoidal magnet.  The
design was optimized for tracking and vertexing capabilities at
pseudorapidities $|\eta|<3$, where $\eta = -\ln(\tan(\theta/2))$ and $\theta$ is the 
polar angle with respect to the proton beam direction ($z$).
Particle energies are measured in three liquid-argon/uranium calorimeters: 
A central calorimeter (CC) covers $|\eta|<1.1$, and two end
calorimeters (EC) extend coverage to $|\eta|<4.2$, each calorimeter housed
in a separate cryostat~\cite{run1det}. 
Central and forward preshower detectors are located just outside 
of the superconducting coil (in front of the calorimetry), and
additional scintillators between the CC and EC cryostats provide
sampling of developing showers for $1.1<|\eta|<1.4$. The muon detection system is outside
the calorimetry and consists of a
layer of tracking detectors and scintillation trigger counters
before 1.8~T iron toroid magnets, followed by two similar layers after
the toroids. The trigger and data acquisition systems are designed 
to accommodate high luminosities.

The dielectron sample is selected from the data by requiring 
two clusters of energy in the electromagnetic (EM) layers at the trigger
level. In the offline selection, two EM clusters
are each required to have transverse momentum $p_{T}>15$
GeV/$c$ and $|\eta|<2.5$. In addition, 
the shower development in the calorimeter and isolation from hadronic activity must
be consistent with that expected of an electron, and at least one of the EM
clusters is required to have an associated track 
to maximize the possibility of having
a $Z$ boson in the event. The electron candidates with matching tracks are
to required to have a ratio of measured energy in the calorimeter
to momentum measured with the tracking system consistent with that expected of an electron. 
The $Z$ candidates are selected by 
requiring a dielectron mass ($m_{ee}$) of $80\ {\rm {GeV}}/c^{2}<m_{ee}<100\ {\rm {GeV}}/c^{2}$.
The $Z+$jet sample is then selected by requiring the presence of at least one reconstructed hadronic 
jet with $p_{T}>20$ GeV/$c$ and $|\eta|<2.5$. 

Jets are reconstructed from calorimeter clusters using a cone algorithm of cone size 
$\Delta {\mathcal R}=\sqrt{{(\Delta\eta)^{2}+(\Delta\phi)^{2}}}=0.5$
in pseudorapidity and azimuth($\phi$). Hadronic jets 
are required to have an associated cluster of tracks (``track jets'').
This requirement reduces background from noise in the calorimeter.
Track jets are found by applying a cone track clustering algorithm of size $\Delta {\mathcal R}=0.5$ with
a seed track of $p_T>1.0\ {\rm GeV}/c$,
to tracks of $p_T>0.5\ {\rm GeV}/c$ that are
close to the primary interaction vertex (whose determination is discussed below). 
A track jet can consist of two or more tracks.

A ``taggable'' jet is 
a calorimeter jet with a matching track jet within $\Delta {\mathcal R}<0.5$.
Applying the taggability criterion to 2,661 jets in 2,219 $Z(ee)$ candidate events 
in the mass $Z$ window yields 1,658 events. 
Based on side bands to the $Z$ mass window, $121\pm 4$ events are 
estimated to be from background sources. The main background is from
multijet production where two jets mimic EM objects,
with one of the objects having an overlapping track that passes the 
track-matching criteria.
The taggability per jet is $(75\pm1)\%$ after background subtraction.

The dimuon sample is defined by the detection of at least one 
muon candidate at the trigger level. In the off\-line selection, two isolated muons 
are required to be of opposite charge, and to have $p_{T}>15$ GeV/$c$ and $|\eta|<2$ with trajectories in
the muon spectrometer matched to tracks in the central-tracking detector.
Muon isolation is based on the transverse component of the muon momentum relative to  
the combined momenta of muon and the closest calorimeter jet in $(\eta,\phi)$ space, and is 
$p_{Trel}>10$ GeV/$c$.
The $Z$ candidates are selected
by requiring a dimuon mass of $65\ {\rm {GeV}}/c^{2}<m_{\mu\mu}<115\ {\rm {GeV}}/c^{2}$.
The $Z$ mass window is larger than in the dielectron channel due to worse momentum resolution
for high $p_T$ muons.
The criteria for reconstructed hadronic jets 
are the same as in the dielectron channel. 
A total of 1,406 events remain after the requirement that there be at 
least one taggable jet. The main background in this channel is from $b\bar{{b}}$ 
production, where both $b$ jets contain muons that satisfy the isolation 
criterion (referred to as $b\bar{b}$ background). 
The isolation efficiencies
of muons from $Z$ and $b\bar{b}$ are expected to be different, since, for the 
latter, a hadronic jet would be expected to be close to the muon. 
By performing fits to dimuon mass spectra, where the background contributes to the
continuum,  samples with different number of isolated
muons are analyzed to measure the isolation efficiencies
and background rates. 
From such analyses, we estimate that the background contribution to the final sample
with two isolated muons is $17.5\pm 4.1$ events.

\begin{figure}
\begin{center}\includegraphics[%
  width=7cm,
  height=7cm,
  keepaspectratio,
  trim=0 40 0 0
  ]{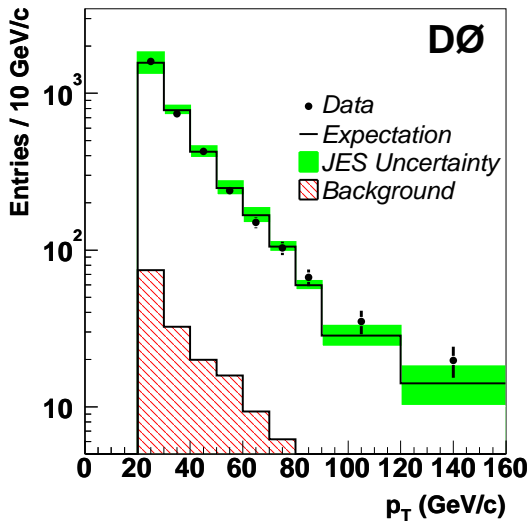}\end{center}

\caption{\label{fig:jetpt}The $p_{T}$ distribution of taggable jets in dielectron
and dimuon channels compared to $Z+$jet \textsc{alpgen} with \textsc{pythia} showering
and full detector simulation (open histogram), and background (multijet for $ee$ channel and $b\bar{b}$ background for $\mu\mu$ channel)
derived from data. The error bars on the data points are statistical. 
The prediction is normalized to the data, as described in the text. }
\end{figure}

Figure \ref{fig:jetpt} shows distributions in transverse momentum of taggable jets for both channels
(points with error bars), compared to a $Z+{\rm jet}$ Monte Carlo (MC)
generated with \textsc{alpgen} \cite{alpgen}, using \textsc{pythia} \cite{pythia} for parton showering
and hadronization. Also shown is a background estimation based on data
obtained from samples that are in the side band for the dielectron channel,
or fail the isolation criterion for the dimuon channel. The background distribution
is normalized to the number of background events estimated in the selected sample.
The simulated signal is then normalized so that the total agrees with the
measurement in Fig.~\ref{fig:jetpt}.
Within the uncertainty of the jet energy scale (JES), indicated by the 
darker shading about the expectation, the shape of the distribution is well-described by the simulation.  

The $b$ quark fragments into a $B$ hadron, which is identified by a 
displaced secondary vertex that is separated from the primary vertex.
The reconstruction of secondary vertices
proceeds in two steps. First, the primary interaction vertex 
is identified, and then additional nearby vertices 
are reconstructed. In the high luminosity environment of the 
Fermilab Tevatron collider, there can be more than one interaction per beam 
crossing, one of which is likely to have triggered the recorded event. The interaction 
region in D\O\ has a root mean-square width of $\approx 25$ cm along $z$, with a
transverse beam size of  $\approx$ 30 $\mu$m.
It is possible to distinguish the main hard-interaction vertex 
from any additional soft interactions because the vertices are normally well-separated 
along $z$. Primary interaction vertices are reconstructed 
in two passes. In the first pass, all tracks present in an 
event are used to find seed vertices using an iterative 
method, where tracks that contribute to a fit to a common vertex 
with a $\chi^{2}/{\rm d.o.f.}$ greater than some chosen threshold are removed.
The fit is repeated until a stable set of seeds is obtained.
The seed vertices are then used in a second pass to fit all tracks 
within a certain distance-of-closest-approach to any seed. This
improves the position resolution on the vertex, since the fit is 
less affected by poorly reconstructed tracks. The $p_{T}$ distribution 
of the associated tracks is then used to select the primary 
interaction vertex (PV).

A $b$-jet tagging algorithm for secondary vertices (SV) is used to identify
heavy-quark jets in the analysis. Tracks that are displaced from the
PV in the transverse plane are used as seeds
to find secondary vertices. First, a fixed-cone jet algorithm of $\Delta {\mathcal R}=0.5$
is used to cluster the tracks to form track-jets. Tracks are required
to have hits in at least two layers of the SMT, $p_{T}>0.5$
GeV$/c$, and be within 0.15~cm in the plane transverse to $z$ and 0.40~cm 
in $z$ relative to the PV. Tracks identified as
arising from $K_{S}^0$ and $\Lambda$ decays or photon conversions are not considered. 
Any pair of tracks within a track-jet with an impact parameter relative to the hard-interaction vertex
(distance of closest approach -- dca -- of a track to a vertex in the plane transverse to the $z$ direction) divided
by their uncertainties ($\sigma_{\rm dca}$), ${\rm dca}/\sigma_{\rm dca}>3$ 
is used as a seed for secondary vertices. Additional tracks are 
attached iteratively to the seed vertices if their $\chi^{2}$-contribution 
to the vertex fit is consistent with originating from the vertex. 
A secondary vertex consists of two or more tracks.
The momentum vector of the SV is defined as the vector sum of track momenta.
Finally, good-quality secondary vertices are selected
based on the decay length (distance between PV and SV), 
collinearity of the vertex momentum with
the direction from PV to SV, and vertex-fit $\chi^{2}$. A jet is
considered $b$ tagged when it is taggable and has at least one secondary
vertex, with a decay-length transverse to the PV ($L_{xy}$) divided by its uncertainty $L_{xy}/\sigma_{xy}>7$,
associated with it. A secondary vertex is
associated to a jet if the opening angle between the direction of the
calorimeter-based jet axis and the momentum vector of the SV
is $\Delta {\mathcal R}<0.5$.

The $b$-tagging efficiency ($\epsilon_{b}$) and the light-flavor
tagging rate ($\epsilon_{L}$) of the $b$-tagging algorithm
are parametrized as functions of jet $p_{T}$ and $\eta$. The parametrization
of $\epsilon_{b}$ is derived from a different data sample using events with 
jets containing muons (muonic jets), which are dominated by $b$ jets, but also have
contributions from light quark jets, gluon jets, and charm jets. 
The $b$-tagging efficiency
is extracted from the heavy-flavor component in this muonic jet sample.
The light-flavor tagging rate
is also derived from data, after compensating for effects of displaced
vertices that do not originate from heavy-flavor decay ($K_{S}^0$,
$\Lambda$, and photon conversions). 
Different types of samples are used to determine 
$\epsilon_{L}$ and $\epsilon_{b}$, and the spreads are taken as systematic uncertainties.

A comparison of inclusive $Z+$jet events, generated with the \textsc{alpgen} leading-order
matrix element and \textsc{pythia} for showering, with inclusive $Z+b$ events generated
with \textsc{pythia}, shows good agreement for jet $p_T$ and $\eta$ distributions.
We therefore use the shapes of $p_T$ and $\eta$ derived from the $Z+$jet data sample to estimate the
expected $b$-tagging efficiency and the light-flavor tagging (``mistag'') rate. 
The average $b$-tagging efficiency 
and mistag rate per jet, averaged over $p_T$ and $\eta$,
are found to be $(32.8\pm1.3)\%$ and $(0.25\pm0.02)\%$, respectively, for the dielectron channel. 
Corresponding values for the dimuon channel are $(33.1\pm1.1)\%$ and $(0.24\pm0.02)\%$.
To obtain the event mistag rate, we take 
into consideration jet multiplicity,
and measure the event mistag rate of 0.28\% (0.27\%) for the dielectron (dimuon) channel.

Since $\epsilon_{b}$ is derived from events with a muon embedded
in a jet, whereas most of the $b$-tagged jets do not contain such
muons, the difference in $b$-tagging efficiencies for hadronic
$b$ jets and muonic $b$ jets is derived from MC, and
the ratio is used to correct $\epsilon_{b}$. 
We cannot at this point derive the charm tagging efficiency ($\epsilon_{c}$) from data, so we rely
on \textsc{pythia} MC to compare $Z\rightarrow b\bar{{b}}$ and
$Z\rightarrow c\bar{{c}}$ samples. We assume that $(\epsilon_{c}/
\epsilon_{b})_{\rm data}=(\epsilon_{c}/\epsilon_{b})_{\rm MC}=0.266\pm0.003$.

The jet taggability, $t_{L}$, is measured using data to be $(75\pm 1)\%$,
while that for $b$ jets, $t_{b}$, is obtained from MC, and scaled
such that $(t_{b})_{\rm data}=(t_{L})_{\rm data}\times(\frac{{t_{b}}}{t_{L}})_{\rm MC}$.
The result is $(t_{b})_{data}=(79.2\pm1.3)\%$ for the dielectron
channel and $(80.7\pm1.1)\%$ for the dimuon channel. We assume that the taggability 
of charm jets is same as $t_b$.

After applying $b$ tagging, 27 $Z(\rightarrow ee)+b$-jet candidate
events are left, with an expected background from the Drell-Yan $ee$ continuum 
and multijet background of $4.2\pm1.4$ events based on the side-band subtraction method. 
In the dimuon channel, 22 events are observed with $5.0\pm1.1$ events from 
$b\bar{b}$ background.

After subtracting the background contributions,
two equations, one before and the other after the requirement of $b$ tagging,
determine the contributions from different flavors in the remaining
events: \begin{eqnarray}
N_{{\rm {before \; b-tag}}} & = & t_{b}N_{b}+t_{c}N_{c}+t_{L}N_{L}\label{eq:Nbefore}\\
N_{{\rm {b-tagged}}} & = & \bar{\epsilon}_{b}t_{b}N_{b}+\bar{\epsilon}_{c}t_{c}N_{c}+\bar
{\epsilon}_{L}t_{L}N_{L},\label{eq:Nafter}\end{eqnarray}
where $N_{b}$, $N_{c}$ and $N_{L}$ are the numbers of events with $b$,
$c$ and light jets, respectively; $t_{i}$ are the taggabilities per event for different
jet types; and the $\bar{\epsilon}_{i}$ are the corresponding mean event-tagging
efficiencies. We assume that the tagging efficiencies per jet, $\bar{\epsilon}_{b}$
and $\bar{\epsilon}_{c}$, are the same as the tagging efficiencies per event.
Equations (\ref{eq:Nbefore}) and (\ref{eq:Nafter}) have three unknowns.
We take the theoretical prediction of $N_{c}=1.69N_{b}$ \cite{campbell}, to provide a  solution of Eqs.
(\ref{eq:Nbefore}) and (\ref{eq:Nafter}) for $N_b$, $N_c$ and $N_L$.

\begin{figure}
\begin{center}\includegraphics[%
  width=4.2cm,
  height=4.5cm,
  trim=0 40 0 0
  ]{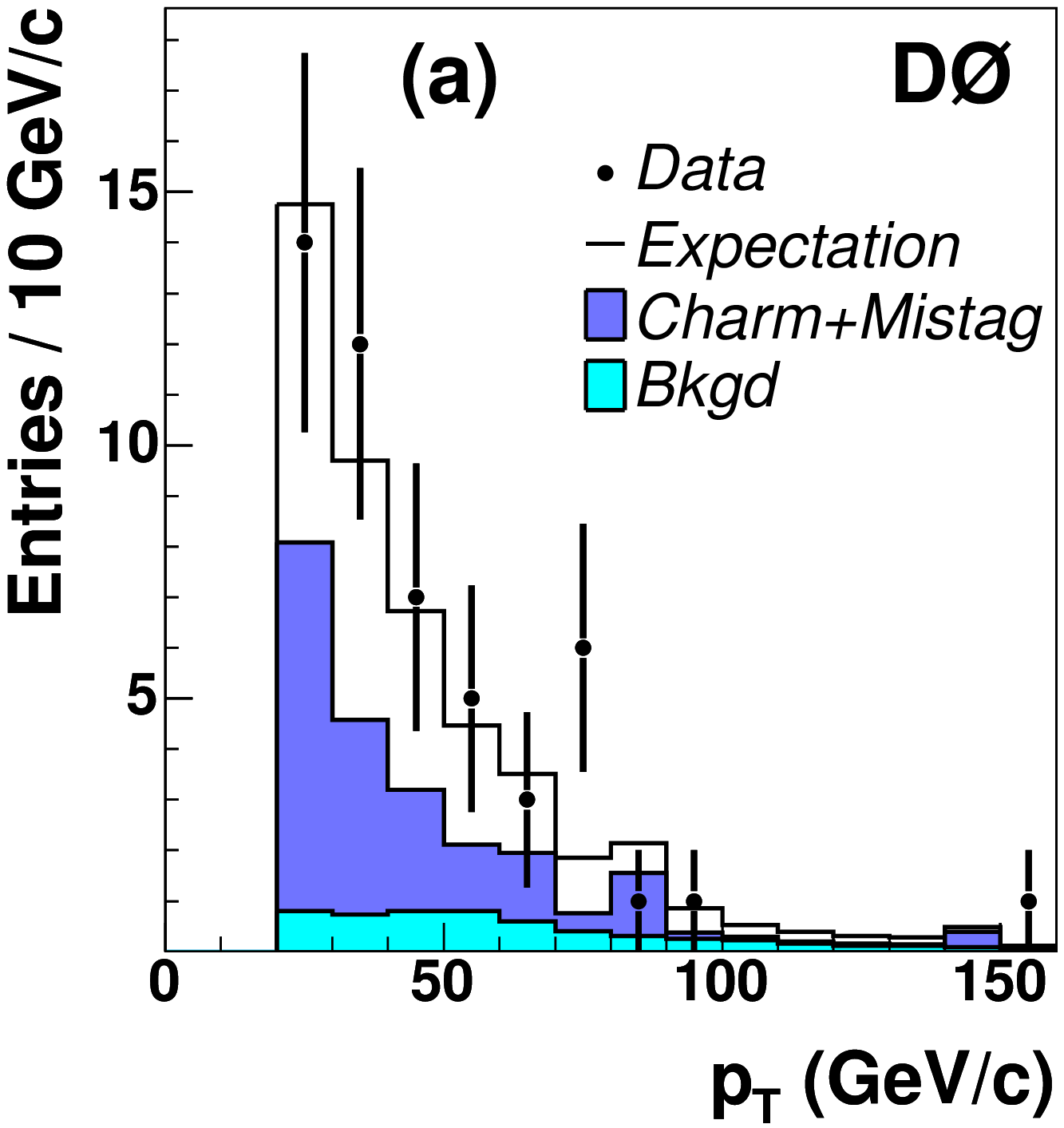}
\includegraphics[%
  width=4.2cm,
  height=4.5cm,
  trim=0 40 0 0
  ]{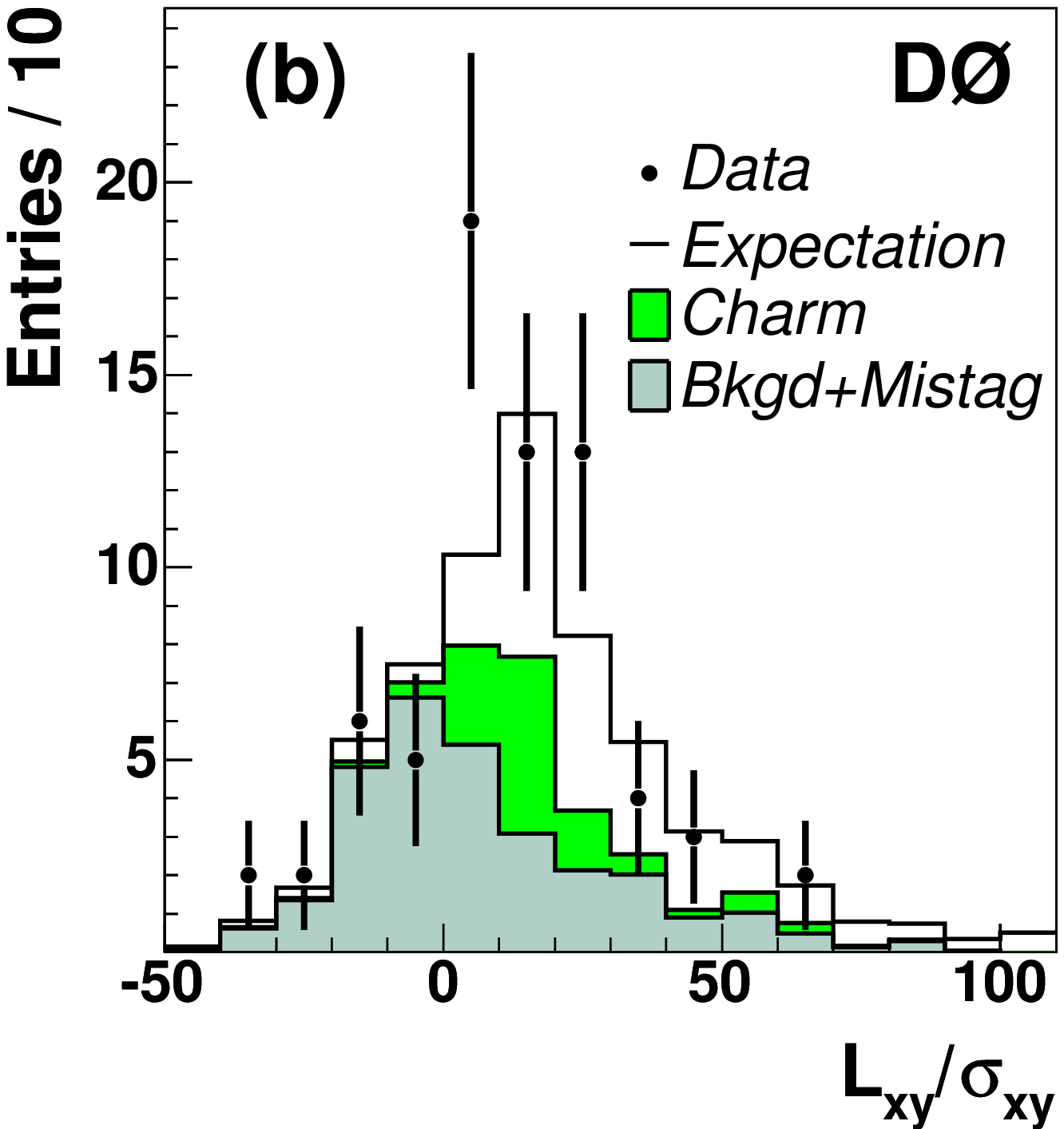}\end{center}

\caption{\label{fig:bjetpt} (a) The $p_T$ spectrum for $b$-tagged jets.
(b) Distribution in decay-length significance of secondary
vertices in the transverse plane, without the requirement on decay-length significance.
All error bars are statistical.}
\end{figure}

The ratio 
$\sigma(p\bar{{p}}\rightarrow Z+b{\rm\ jet})/\sigma(p\bar{{p}}\rightarrow Z{\rm+jet})
=N_{b}/(N_{b}+N_{c}+N_{L})$
is found to be $0.026\pm0.007$ for the dielectron channel and $0.020\pm0.005$
for the dimuon channel, where the errors are purely statistical.
The combined ratio, using the statistical weighting of the number of observed
$Z+$jet candidates, is $0.023\pm0.004$. The shape of the  $p_{T}$ spectrum for 
$b$-tagged jets and the significance of decay lengths of secondary vertices
are compared to the sum of background and 
$Z+b$ MC in Fig. \ref{fig:bjetpt}. The contribution of each component
is given by the solution to Eqs. (\ref{eq:Nbefore}) and (\ref{eq:Nafter}).
The distribution of the decay-length significance 
for secondary vertices shows clear evidence 
for a heavy-flavor component in the $b$-tagged candidate events.

%
%
%

%

%
%
%

\begin{table}
\caption{Systematic uncertainties for the combined ratio of cross sections, showing the impact
of $\pm 1$ standard deviation changes in contributions\label{tab:syst}.}
\begin{center}\begin{tabular}{lcc}
\hline
Source& Upward (\%)& Downward (\%)\tabularnewline
\hline
Jet energy scale& 5.7& 6.7\tabularnewline
Background estimate& 5.6& 5.3\tabularnewline
$Z+(Q\bar{Q})$ & 0.0 & 5.5 \tabularnewline
Mistag rate& 3.5& 3.3\tabularnewline
$b/c$ tagging efficiency& 3.1& 3.2\tabularnewline
Taggability& 1.9& 3.7\tabularnewline
Correction for hadronic jet& 2.0& 1.7\tabularnewline
Jet reconstruction efficiency& 1.8& 1.8\tabularnewline
$\sigma(Z+c)/\sigma(Z+b)$& 2.8& 2.9\tabularnewline
\hline
\hline
Total (added in quadrature)& 10.2& 12.3\tabularnewline
\hline
\end{tabular}\end{center}
\end{table}

Sources of systematic uncertainty in the ratio include:
\begin{list}{\roman{enumi})}{\usecounter{enumi}\setlength{\itemsep}{0pt}}
\item Jet energy scale. The JES is varied  within its uncertainty.
The JES for hadronic $b$ jets is
assumed to be the same as that for light-flavored jets, 
whereas in MC some differences are observed, and this effect is included as
part of the JES uncertainty.
\item Different methods of estimating background. The background is varied by its measured uncertainty
and the ratio is recalculated.
\item Jets that contain a $b\bar{b}$ or $c\bar{c}$ pair from gluon splitting.
These jets have a higher tagging probability. 
The expected contribution is taken from 
theory \cite{campbell}, and the relative increase in $b(c)$-tagging efficiency is 
estimated from MC.  This is labeled as $Z+(Q\bar{Q})$ in Table \ref{tab:syst}.
\item Mistag rate for light jets, which depends on the type of jet sample.
Using events collected from hadronic jet triggers,
the light-jet tagging efficiency is measured to be $0.23\%$, and for a 
sample of events with an enhanced EM fraction and small imbalance in
overall $p_T$, this is $0.26\%$. A tagging efficiency of $0.25\%$
per jet (or $0.28\%$ per event) is obtained for the combined data. 
\item Uncertainty in tagging efficiency for $b$ and $c$ jets is
obtained by varying the efficiency by a $\pm1$ standard  
deviation, assuming complete correlation in the ratio of 
extracted cross sections. Also, for $c$ jets, there is 
additional uncertainty from the $\epsilon_{c}/\epsilon_{b}$ ratio obtained
from MC.
$\bar{\epsilon}_{L}$ is varied, as above, to estimate this effect.
\item A small difference observed in $t_{b}/t_{L}$ for different MC samples
of $Z+b$ jet$/Z+$light jet, and $Z\rightarrow b\bar{{b}}/Z\rightarrow q\bar{{q}}$ is taken into account.
\item Differences in  tagging efficiency between hadronic jets and those containing muons.
The $b$-tagging efficiency is measured in data using 
muonic jets. The tagging efficiency for hadronic jets is estimated to be $86\%$ 
of that of muonic jets, as derived from $Z\rightarrow b\bar{{b}}$ MC. 
The same ratio in $Z+b\bar{{b}}$ MC is measured to be $84\%$, and the difference
of $2\%$ is taken as a systematic uncertainty.
\item Different $p_{T}$-dependence in jet reconstruction for
light, $b$, and $c$ jets, measured using MC samples, is accounted for as a systematic uncertainty.
\item Uncertainty from theory for the ratio $\sigma(Z+c~{\rm\ jet})/\sigma(Z+b~{\rm\ jet})=N_c/N_b$ is estimated as 
9.5\% \cite{campbell}.
\end{list}

The effects of systematic uncertainties on the combined measurement
are listed in Table \ref{tab:syst}. All these uncertainties
are assumed to be completely correlated for the two channels, except for
that due to background estimation. Folding these  
uncertainties together, yields a ratio of
$0.023\pm0.004({\rm {stat}})^{+0.002}_{-0.003}({\rm {syst})}$. 
This measurement is in good agreement with the next-to-leading order (NLO) prediction of 
$0.018\pm 0.004$ \cite{campbell2}. 

In summary, we have presented the first inclusive measurement of $b$-jet production in association 
with $Z$ bosons at the Fermilab Tevatron collider, which is a background to the standard-model Higgs searches
in the $ZH$ production channel. The measurement is in agreement with the NLO
calculations and can be used to constrain the $b$-quark density of proton.

%
We thank the staffs at Fermilab and collaborating institutions, 
and acknowledge support from the 
Department of Energy and National Science Foundation (USA),  
Commissariat  \` a l'Energie Atomique and 
CNRS/Institut National de Physique Nucl\'eaire et 
de Physique des Particules (France), 
Ministry of Education and Science, Agency for Atomic 
   Energy and RF President Grants Program (Russia),
CAPES, CNPq, FAPERJ, FAPESP and FUNDUNESP (Brazil),
Departments of Atomic Energy and Science and Technology (India),
Colciencias (Colombia),
CONACyT (Mexico),
KRF (Korea),
CONICET and UBACyT (Argentina),
The Foundation for Fundamental Research on Matter (The Netherlands),
PPARC (United Kingdom),
Ministry of Education (Czech Republic),
Natural Sciences and Engineering Research Council and 
WestGrid Project (Canada),
BMBF and DFG (Germany),
A.P.~Sloan Foundation,
Research Corporation,
Texas Advanced Research Program,
and the Alexander von Humboldt Foundation.
%

\end{document}